\documentclass[preprint,12pt]{elsarticle}

\usepackage{amssymb}
\usepackage{algorithm}
\usepackage{algpseudocode}

\usepackage{amsmath}

\usepackage{xcolor}
\usepackage{appendix} 
\usepackage{subcaption}

\journal{Chaos Solitons and fractals}

\begin{document}

\begin{frontmatter}



\title{A Grand-Canonical Solution to a Class of Random Optimization Problems} 

\author[1,2]{Izat B. Baybusinov}
\author[3]{Enrico Maria Fenoaltea\corref{cor1}}
\author[1]{Zhen Han}
\author[1]{Yi-Cheng Zhang}

\cortext[cor1]{Corresponding author: enricofenoaltea@hotmail.it}

\affiliation[1]{
organization={
International Research center for Complexity Sciences, Hangzhou international innovation institute of Beihang University},
city={Hangzhou},
state={China},
postcode={311115}}

\affiliation[2]{
organization={ Physics Department, University of Fribourg},
addressline={Chemin du Musée 3},
postcode={1700},
city={Fribourg},
country={Switzerland}}
\affiliation[3]{
organization={Universitat de Barcelona Institute of Complex Systems (UBICS), Universitat de Barcelona},
postcode={08028},
city={Barcelona},
country={Spain}}




\begin{abstract}
We introduce a general framework to solve a class of combinatorial optimization problems, including the matching problem, the Traveling Salesman Problem, and also the minimum weight k-factor problem. By reformulating these problems as an arrangement model, we recast the optimization task into a grand-canonical ensemble, where chemical potentials are used to relax strict topological constraints. The analytical solution found can serve as a polynomial-time algorithm to compute an approximate minimum cost for arbitrary k and link-weight distributions. Our framework is complementary to existing approaches and reveals new connections between combinatorial optimization and the statistical physics of disordered systems.
\end{abstract}


\begin{highlights}
\item We recast optimization problems, such as the matching problem and the traveling salesman problem (TSP) as an arrangement model
\item We solve them within the grand canonical ensemble
\item Our framework offers an intuitive approach to analytically solve hard optimization problems
\item It also provides a polynomial-time algorithm to approximate solutions to combinatorial problems
\end{highlights}


\end{frontmatter}



\section{Introduction}
Imagine having to arrange N people around a table for a dinner. Each guest has its own preferences, people they would love to sit next to and others they would rather avoid. If the goal is to maximize happiness, what seems like a simple social scenario quickly becomes a challenging optimization problem. 
We define this kind of scenario as an arrangement problem. This belongs to a broader class of combinatorial optimization problems, such as the well-known Travelling Salesman Problem (TSP) \cite{gavish1978travelling}. In fact, when individuals are seated around a circular table, our arrangement problem can be shown to be equivalent to the TSP.


Other problems within the same class include the matching problem, where each agent is paired with only one other neighbor \cite{fenoaltea2021stable}, and the more general minimal weight k-factor (or monopartite k-assignment) problem, where every agent must have exactly
k neighbors \cite{dey2024random, sicuro2021planted}. When the weights in these problems are drawn randomly, they fall under the category of random optimization problems \cite{lueker1981optimization, mertens2000random}.
This type of research has attracted significant interest across fields, from physics and mathematics \cite{nishimori2001statistical, korte2011combinatorial} to computer science and biology \cite{papadimitriou1998combinatorial, naseri2020application}, and in particular within the interdisciplinary network-science community, as many network models can be reformulated as constrained optimization problems with many real-world applications \cite{park2004statistical}. 

While simple to state, these optimization problems are notoriously difficult to solve. The difficulties primarily stem from the absence of translational symmetry and the presence of frustrations, which prevent straightforward local optimization methods from identifying the true ground state \cite{moore1987zero}.
In mathematical programming formulation, one seeks to minimize a cost function subject to
a set of constraints, a setting naturally treated using Lagrangian duality
\cite{bertsekas1997nonlinear}. When the constraints are linear, as in the matching
problem, one can apply a Lagrangian relaxation  introducing multipliers that
penalize violations of the degree constraints. This leads to an optimization problem
whose linear-programming relaxation is convex and, in this
case, exactly solvable \cite{boyd2004convex}. By contrast, in problems such as the TSP or map coloring problems, which involve
global and nonlinear structural constraints, Lagrangian relaxations yield only
approximate solutions \cite{held1970traveling, guignard2003lagrangean}. For such
problems, approaches based on belief-propagation algorithms have emerged as
efficient numerical approximation methods~\cite{hartmann2006phase,
feige2006complete}.

From an analytical point of view, statistical physics emerged as a powerful tool for addressing these problems. The first steps in bridging the analytical challenges of combinatorial optimization with statistical-mechanics methods were taken by Parisi and Mézard in their pioneering works~\cite{mezard1985replicas, mezard1986mean, mezard1987spin}.
The connection between statistical mechanics and optimization was later further developed within network theory~\cite{park2004statistical}, for example through maximum-entropy network models used for network reconstruction~\cite{squartini2017maximum} and validation~\cite{bruno2023inferring}. In these approaches, one seeks maximum-entropy ensembles of networks subject to structural constraints enforced by Lagrange multipliers, such as prescribed degrees, edge weights, or reciprocity~\cite{squartini2011analytical, cimini2019statistical}, in direct analogy with the constrained optimization problems discussed above. One can also extend these problems on networks to incomplete optimization \cite{buffa2025maximum}, introducing a temperature-like parameter that interpolates between the cost-efficient (zero-temperature) limit and the random (infinite-temperature) regime, enabling controlled exploration of near-optimal configurations.

While statistical-mechanics approaches have been remarkably successful in producing exact analytical results for several optimization problems, they remain conceptually challenging. As a consequence, a substantial mathematical literature has developed to formalize and rigorously justify the insights obtained from statistical physics. For instance, Parisi’s solution of the matching problem via the replica method~\cite{mezard1985replicas} was rigorously established only fifteen years later by Aldous~\cite{aldous2001zeta, nair2005proofs} using an infinite-tree limit argument.

In this paper, we address random optimization problems from a new perspective and derive the average ground-state energy in a more intuitive and accessible manner. Specifically, we introduce a general framework by reformulating three classical random optimization problems (matching, TSP, and minimum weight k-factor) as an arrangement model, in which a set of agents must be arranged on a network. This formulation naturally leads to a grand-canonical ensemble approach, simplifying their analytical treatment. We recover the classical solutions for both the matching problem and the TSP found in \cite{aldous2001zeta} which we identify as a local Lagrangian relaxation where only local properties of the network are considered. We show how our framework allows for an intuitive derivation of the average ground-state energy with a polynomial approximation algorithm to solve this class of combinatorial optimization problems for a given realization of weights. Finally, we discuss the similarities, differences, and complementary insights that such a grand-canonical formulation offers relative to existing statistical-physics-based approaches in the literature.

\section{The general framework}
We consider a network with $N$ nodes and an adjacency matrix $\mathbf{A}$, where $A_{ab} = 1$ if node $a$ is connected to node $b$, and $A_{ab} = 0$ otherwise. Each agent has to occupy a node in the network, effectively forming a seating arrangement. This arrangement can be represented by a permutation matrix $\mathbf{X}$, where
$X_{ia} = 1$ if agent $i$ is seated at node $a$, and $X_{ia} = 0$ otherwise.
Since each agent can occupy only one seat, the permutation matrix must satisfy $\sum_{a=1}^{N} X_{ia}= 1$ for any $i$.
This representation defines the configuration space of all possible seating arrangements.

The connection cost of an agent $i$ depends on their neighbors in the network. Each agent has a preference list for other agents, which, in the statistical physics framework, is encoded as an energy matrix $\mathbf{\epsilon}$. The element $\epsilon_{ij} \in [0,1]$ represents the interaction energy (or cost) for agent $i$ sitting next to agent $j$. Given a network topology $\mathbf{A}$ and an energy matrix $\mathbf{\epsilon}$, the energy $h_i[\vec{X}_{i}]$ of agent $i$ in a given configuration $\mathbf{X}$ is given by
\begin{equation}
    h_i[\vec{X}_{i}] = \sum_{a=1}^N \sum_{j=1}^N \epsilon_{ij} X_{ia}\left( \sum_{b=1}^N X_{j b} A_{ab} \right) .
\end{equation}

The global energy of the system, i.e. the Hamiltonian, is thus the sum of these terms:
\begin{equation}\label{H}
    H[\mathbf{X}] = \frac{1}{2} \sum_{i,j=1}^N (\epsilon_{ij}+\epsilon_{ji} )\left(\sum_{a,b=1}^N X_{ia}X_{jb} A_{ab}\right).
\end{equation}


For networks with specific topologies, the introduced framework reduces to a class of well-known constraint optimization problems. For example, if we impose that each node has exactly one neighbors, i.e., $\sum_{b} A_{ab} =\sum_{i=1}^{N} X_{ia}= 1 \, \forall a$, the problem of minimizing Eq.\ref{H} becomes equivalent to the classical matching problem \cite{kuhn1955hungarian, mezard1985replicas}. If we require the network to be a chain, i.e., $\sum_{i=1}^{N} X_{ia}= 1$ and $\sum_{b}A_{ab}=2 \, \forall a$ with the connectivity constraint (the network must consist of only one- non disconnected chain), the Hamiltonian to minimize becomes
\begin{equation}
H[\mathbf{X}] = \frac{1}{2} \sum_{i,j=1}^{N} (\epsilon_{ij}+\epsilon_{ji}) \left(\sum_{a=1}^{N} X_{ia}X_{j(a+1)}\right),
\label{H_min}
\end{equation}
which corresponds to the TSP \cite{gavish1978travelling, vannimenus1984statistical}. Note that this formulation of the TSP is equivalent to the seating problem introduced in \cite{fenoaltea2021stable}, that is a generalization of the TSP with a non-symmetric energy matrix. Instead, if the connectivity of a single chain is relaxed, the minimization of the Hamiltonian reduces to the minimum weight 2-factor problem \cite{matsiy2016fast, karp1979patching}.

In the In the following, we focus on solving the
constrained optimization problems embedded in a monopartite and unweighted network topology, as those described above. However it is worth to mention that, if one allows the network topology to be bipartite, so that the
row and column sums of $\mathbf{A}$ may differ, i.e.,
$\sum_{b} A_{ab} = k_a \,\, \forall a$ and $\sum_{a} A_{ab} = k_b \,\, \forall b$, the problem reduces to the discrete optimal transport
problem \cite{villani2008optimal}. Also, if $k_a$ and $k_b$ indicate not the prescribed row or column sums but rather their maximum allowed values, one recovers the general assignment problem \cite{burkard2012assignment,fenoaltea2021stable}. If we consider a fully connected weighted network, where $A_{ab}$ is not
binary but can take arbitrary values, the goal becomes to minimize the interaction cost between agents entangled with the distance between seats. In this way, the arrangement model in Eq.~\eqref{H} becomes equivalent to the quadratic assignment problem \cite{loiola2007survey}, that is also equivalent to the nestedness maximization problem, as shown
in \cite{mariani2024ranking}, whose solution has implications for the stability and dynamics of ecological and economic communities.

\section{Results}
To find the ground state of our arrangement model, we first rewrite the Hamiltonian defined in Eq.~\eqref{H} as
\begin{equation}\label{H_meanfield}
\tilde{H}[\mathbf{V}] = \sum_{i<j} E_{ij} V_{ij},
\end{equation}
where $\mathbf{V}$ represents the underlying agent network corresponding to a given configuration $\mathbf{X}$, with elements defined as
\begin{equation}
V_{ij} = \sum_{a=1}^N \sum_{b=1}^N X_{ia} X_{jb} A_{ab},
\end{equation}
and $E_{ij}$ denotes the weight associated with the connection between agents $i$ and $j$. For a general k-factor problem the following constraints must be satisfied:\footnote{One may additionally impose the avoidance of self-loops by setting $V_{ii} = 0$. For generality, we also allow self-loops. However, in practice the avoidance of self-loops is usually encoded directly in the cost matrix by setting $E_{ii} = \infty$.}
\begin{equation}\label{condition}
V_{ij} = V_{ji}, \quad V_{ij} \in {0,1}, \quad \sum_{j} V_{ij} = k.
\end{equation}
Note that these conditions must also hold for the TSP ($k=2$), regardless of the connectivity constraint. Indeed, the main approximation in the following derivation of the ground state of the TSP consists in considering only the local properties of $\mathbf{V}$, while neglecting higher-order terms—namely, higher-order topological constraints. This is justified by the fact that, in the thermodynamic limit, the energy of the minimum weight 2-factor and the TSP converge \cite{frieze2004random}. 

In general, we solve our arrangement problem through the grand canonical ensemble by introducing $N$ chemical potentials ${\mu_i}_{i=1,...,N}$. The partition function is then defined as:
\begin{equation}\label{partition}
Z({\mu_i}, \beta) = \prod_{i,j}\sum_{V_{ij}=0}^1 \exp\left\{-\beta \tilde{H}[\mathbf{V}] - \beta \sum_{i,j=1}^N \mu_i V_{ij}\right\},
\end{equation}
with
\begin{equation}\label{relax}
-\frac{1}{\beta}\frac{\partial}{\partial \mu_i} \log Z = \overline{\sum_{j=1}^N V_{ij}} = k, \quad \text{for } i=1,\ldots,N.
\end{equation}
Equation~\eqref{partition} defines the grand canonical partition function, and Eq.\eqref{relax} enforces the constraint on the average node degree of the network $\mathbf{V}$. 

This approach conveniently avoids the complexities of replica or cavity methods, enabling explicit computation of the partition function:

\begin{equation}\label{partition2}
   Z(\{\mu_i\}, \beta) = \prod_{i,j} \left( 1 + \exp\left\{-\frac{\beta}{2}\left(E_{ij} - \mu_i - \mu_j\right) \right\} \right).
\end{equation}

This expression is a standard Fermi-type partition function and Eq.~\eqref{relax} becomes
\begin{equation}\label{chemical_potentials}
    \sum_{j=1}^N \frac{1}{1 + \exp\left\{\frac{\beta}{2}\left(E_{ij} - \mu_i - \mu_j\right)\right\}} = k,
\end{equation}
which is a system of $N$ equations that, given a set of random energies $\epsilon_{ij}$ (or random link weight $E_{ij}$), can be solved numerically to find ${\mu_i}_{i=1,...,N}$. Once the chemical potentials are computed, we can plug them into the partition function and calculate the average energy $\overline{\tilde{H}[\mathbf{V}]}$.  



To obtain the ground-state energy 
$U_{min}$, we take the zero-temperature limit, yielding:
\begin{equation} \label{eq:min_energy}
    U_{min} =\lim_{\beta\rightarrow\infty} \frac{1}{2}\sum_{ij} \frac{ E_{ij}}{{1+e^{\frac{\beta}{2}\left(E_{ij} - \mu_i - \mu_j\right) }}}.  
\end{equation} 

This equation is valid for a given realization of the energy matrix $\mathbf{\epsilon}$ and the chemical potentials depend on the energies. The configurational average of the ground-state energy $\langle U_{min}\rangle_{\epsilon}$, i.e the average over all possible realizations of the energy matrix, can then be written as follows:
\begin{equation}\label{eq_media}
\begin{split}  
    \langle U_{min}\rangle_{\epsilon} &=  \frac{1}{2}\sum_{ij} \int d\mathbf{E} \rho(\mathbf{E}) E_{ij} \theta(\mu_i +\mu_j -E_{ij} ) =\\
    &=\frac{N^2}{2} \int d E_{12} \rho(E_{12})\,E_{12} \,\text{prob}(\mu_1+\mu_2 > E_{12} )  
\end{split}
\end{equation}
Here $\rho(\mathbf{E}) = \prod_{ij} \rho(E_{ij})$ denotes the joint probability
distribution of the energies, where we consider connection energies to be
i.i.d., as is standard in random optimization problems. In the last step of
Eq.~\ref{eq_media}, we integrate over all energy variables not appearing in the integrand, i.e., all $(a,b)\neq(i,j)$. This results in the term
$\text{prob}(\mu_1+\mu_2 > E_{12})$, which arises from the presence of the
Heaviside theta function. With this formulation we retrieve the minimum energy equation derived by Aldous in \cite{aldous2001zeta}. 

\begin{figure}[p]
\centering

\begin{subfigure}{\textwidth}
    \caption{$k=1$}  
    \centering
    \includegraphics[scale=1]{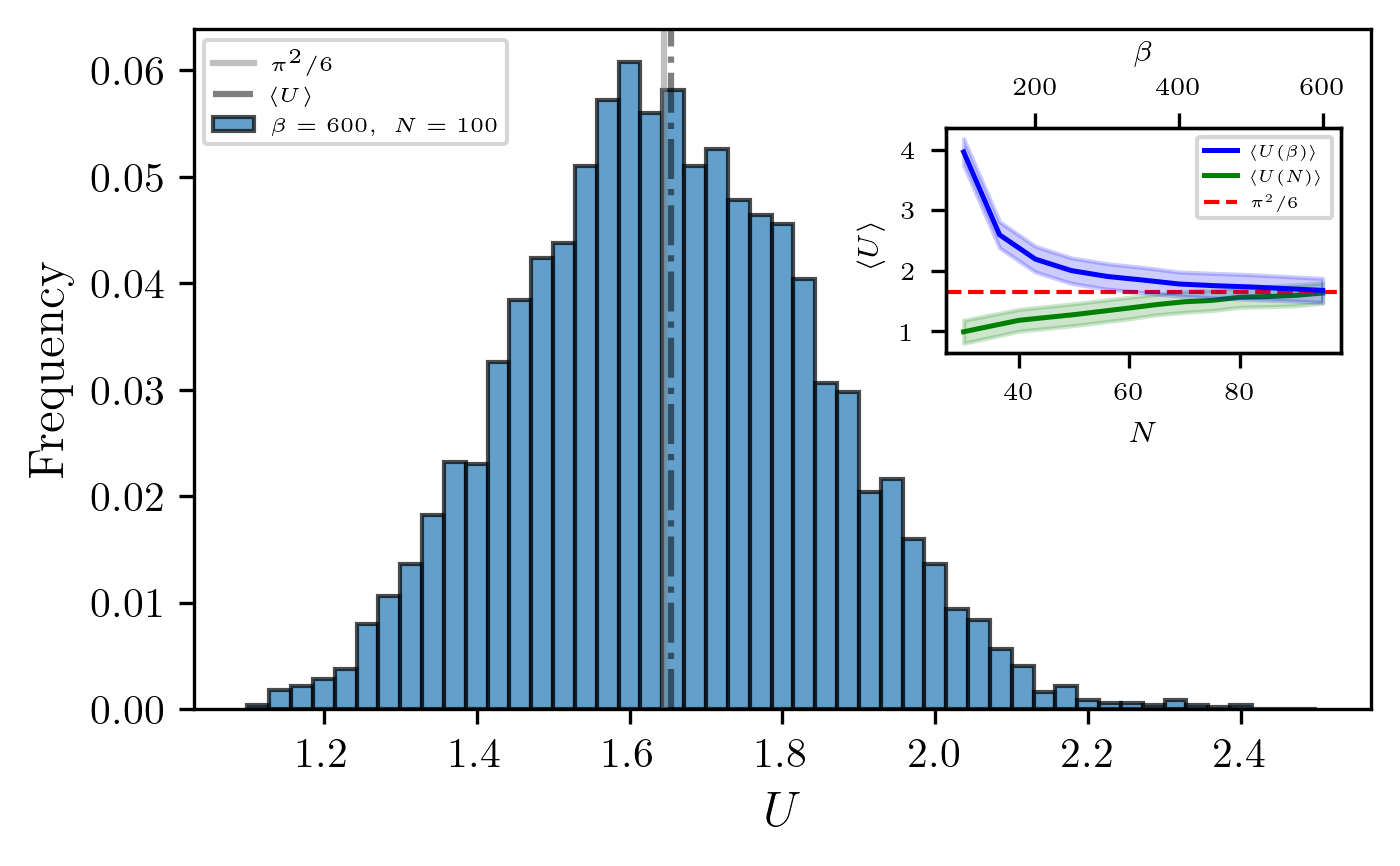}
    \label{fig:E_a}
\end{subfigure}

\begin{subfigure}{\textwidth}
    \caption{$k=2$}  
    \centering
    \includegraphics[scale=1]{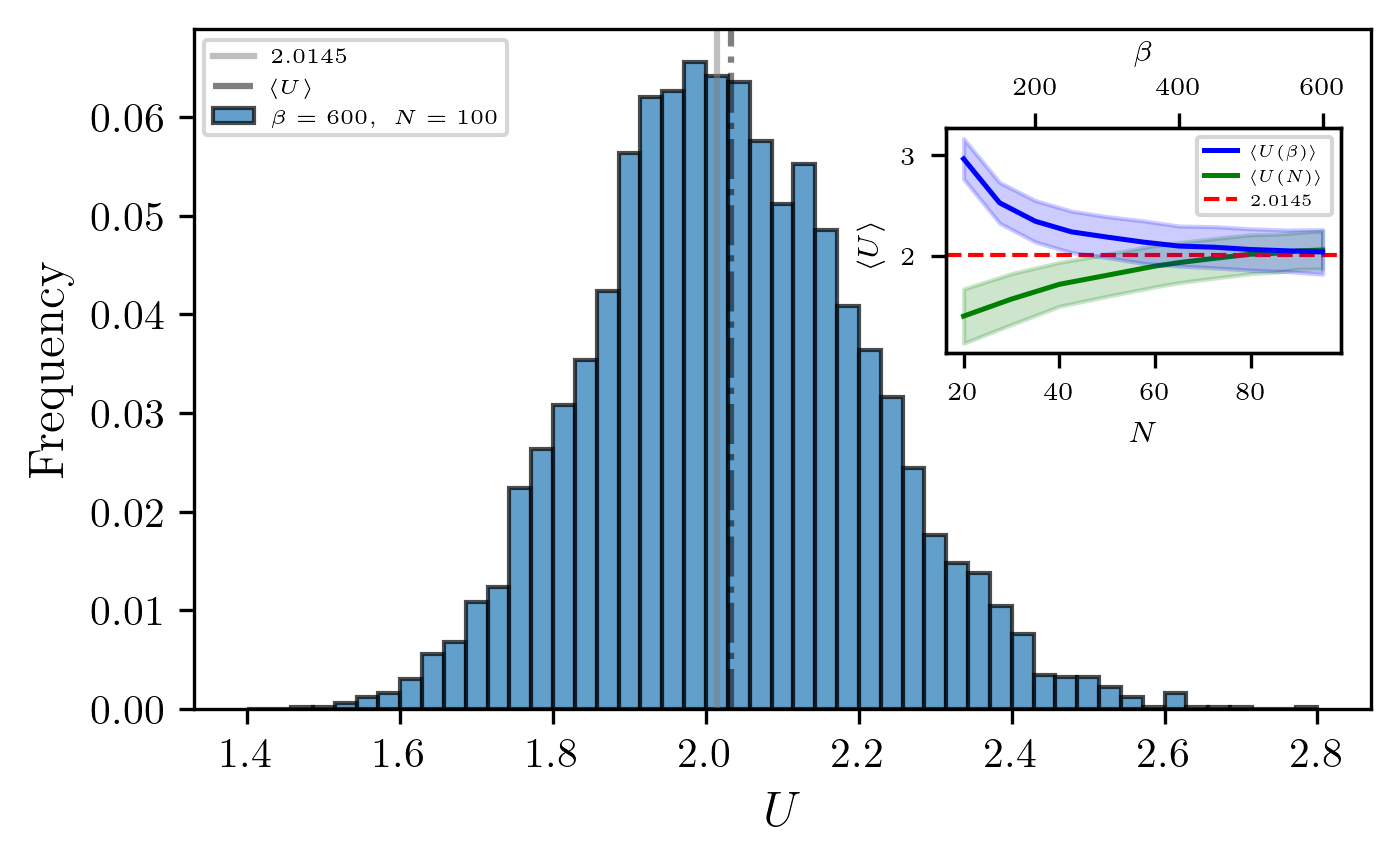}
    \label{fig:E_b}
\end{subfigure}

\caption{Distribution of energies computed from Eq.~\eqref{eq:min_energy} over 5000 realizations for a system of $N = 100$ sites with $\beta = 600$. The top panel (a) corresponds to the matching problem ($k = 1$), and the bottom panel (b) to the minimum weight 2-factor problem ($k = 2$). The solid gray lines represent the classical solutions derived in \cite{mezard1985replicas, mezard1986replica}, while the black dashed line denotes the numerical sample average. The insets show the average energy as a function of $\beta$ and $N$, where the shaded region indicates the standard deviation and the dashed line shows the theoretical values. Note that the averages shown in the insets are necessarily computed at large but finite \(N\) and \(\beta\), respectively. The theoretical value is obtained in the thermodynamic (\(N \to \infty\)) and zero-temperature (\(\beta \to \infty\)) limits. As a result, convergence in the figure is not exact: \(\langle U(\beta) \rangle\) should cross the theoretical value from above as \(\beta\) increases, while \(\langle U(N) \rangle\) should cross it from below as \(N\) increases.
}
\label{fig:E}
\end{figure}

It is interesting to study how the above solution scales with $N$. Since we are working in the grand canonical ensemble, each term in Eq.~\eqref{eq:min_energy} becomes independent when averaging over all possible energy realizations. Thus, in the limit $\beta \to \infty$, each addend in the sum can be replaced by its minimum value (up to a multiplicative constant). Thus we find
\begin{equation}\label{scaling}
\langle U_{\text{min}}\rangle_{\epsilon} \approx N \, \overline{\min_j E_{ij}}.
\end{equation}
This naturally gives a polynomial-time algorithm ($\sim N^2$) to approximate the configurational average of our arrangement model's ground state energy. Instead, for a single instance of the energy matrix, one can use Eqs.\eqref{chemical_potentials} and \eqref{eq:min_energy} to compute in polynomial time an estimate of the minimum cost (see Appendix A for details).

Note that the addends in Eq. \eqref{chemical_potentials} forms a non-integer doubly stochastic matrix, i.e., that satisfies the specified constraints over both rows and columns. By the Birkhoff–von Neumann theorem \cite{schrijver2003combinatorial}, such matrix can be expressed as a convex combination of permutation matrices. In \cite{kosowsky1994invisible}, where a system of equations related to \eqref{chemical_potentials} is derived (which we discuss at the end of this section), it is rigorously shown that this property ensures that the energy defined in Eq.~\eqref{eq:min_energy} converges, as \(\beta \to \infty\), to the exact ground-state energy of the matching problem for \(k = 1\), provided it has a unique solution. If this is not the case the case, i.e., the optimal solution is degenerate, we do not have a formal proof that Eqs.~\eqref{chemical_potentials} converge to the optimal configuration. In trivially degenerate cases, such as when two or more rows of the energy matrix $E_{ij}$ are identical, different initializations of the chemical potentials may lead to different optimal solutions. Nevertheless, in the random case, where the energies are iid continuous variables, degeneracy occurs with probability zero \cite{linusson2004proof,aldous2001zeta}. By virtue of the Tutte f-factor theorem \cite{anstee1985algorithmic}, which states that any k-assignment can be mapped to a matching, the above result extends more generally to the minimum-weight k-factor problem. For the TSP, this is not true in general, because the additional constraint requiring the network to form a single Hamiltonian cycle invalidates the applicability of the Birkhoff–von Neumann theorem. However the same convergence holds in the thermodynamic limit \(N \to \infty\), since in this limit, the ground-state energy of the minimum weight 2-factor problem coincides with that of the TSP \cite{frieze2004random}. These results confirm the validity of our grand canonical framework.

To illustrate, we numerically implement the above framework for  $k=1$ and $k=2$ with a uniform distributed and symmetric energy matrix, and we finds the classical results obtained in \cite{mezard1985replicas, mezard1986replica, fenoaltea2021stable} for the average energy of the matching problem and TSP.

Also, our approach allows to derive the probability that a link between agents $i$ and $j$ exists in the optimal configuration for a given $\beta$, from which we can compute the most probable solution to the arrangement problem from a given realization of costs. In particular, from \eqref{partition} and \eqref{partition2}, the expected value of $V_{ij}$ (or the probability of $V_{ij}$ to be one) can be written as:
\begin{equation}\label{probV}
\overline{ V_{ij}}=\frac{1}{\beta}\frac{\partial \log Z}{\partial E_{ij}}=\frac{1}{{1+e^{\frac{\beta}{2}\left(E_{ij} - \mu_i - \mu_j\right) }}}. 
\end{equation}
When $\beta \to \infty$, this becomes a Heaviside function, $\overline{V_{ij}} \to \theta(E_{ij} - \mu_i - \mu_j)$. Due to the double-stochasticity properties mentioned above, one can show that in this limit, both for the matching problem and for the general minimum-weight $k$-factor problem, $\overline{V_{ij}} \to V^{*}_{ij}$,
where $V^{*}_{ij}$ is the optimal network between agents that minimizes the energy in Eq. \eqref{H_meanfield}. In Appendix B we illustrate this with simple matching examples. In the case of the TSP (or equivalently the sitting problem), such convergence does not generally hold, again because the Birkhoff–von Neumann theorem is not applicable due to the additional constraint that the solution must form a unique Hamiltonian cycle. However, we can use the elements $V_{ij} > 0$ at finite temperature to develop a greedy algorithm that efficiently approximates the solution of the TSP. Indeed, in our grand-canonical formulation, for
finite $\beta$ the elements in Eq.~\eqref{probV} satisfy the constraint that the sum over rows and columns equals $k = 2$ on average. Hence, among all the possible configurations there exist some that correspond to unique Hamiltonian cycles. By following the most probable $V_{ij}$ to be constrained on a cycle, with a sufficiently low temperature we
can approach the cycle with minimum energy. In Appendix B, we present the algorithm in detail together with results on benchmark TSP distance matrices.

We now discuss how the results presented above complement and provide new insights compared to existing approaches in the literature. As previously mentioned, a system of equations closely related to Eq.~\eqref{chemical_potentials} has been derived using a saddle-point approximation in \cite{kosowsky1994invisible} and more recently in \cite{koehl2021fast, koehl2024general}, in the context of the bipartite assignment problem (see Appendix C for details of the corresponding derivation in our monopartite setting). The saddle-point approximation requires large $N$, and assumes that the dominant contribution to the partition function comes from a single configuration. This is exact only in the zero-temperature limit. This approximation is necessary because the partition function derived in these approaches is not analytically tractable and must be justified \textit{a posteriori} by proving that the saddle point is unique and corresponds to the minimum-energy state for $\beta \to \infty$.

In our case, the grand canonical formulation is adopted \textit{a priori} and, together with the arrangement model formulation, provides a more intuitive interpretation grounded on standard statistical mechanics. Specifically, in the grand canonical ensemble, our arrangement model allows multiple agents to occupy the same position in the network with nonzero probability, while ensuring that the average occupancy per site over many realizations is fixed to one. In this setting, agents are not restricted to a single position but are free to explore the network, with their freedom tuned by the temperature (see appendix D for a further discussion). In our formalism we do not need large $N$ and we are interested in the most probable configuration for any $N$ and $\beta$. In the $N \to \infty$ limit the two approaches coincide as is the case in thermodynamics.

\begin{figure}[p]
\centering

\begin{subfigure}{\textwidth}
    \caption{$k=1$}  
    \centering
    \includegraphics[scale=0.5]{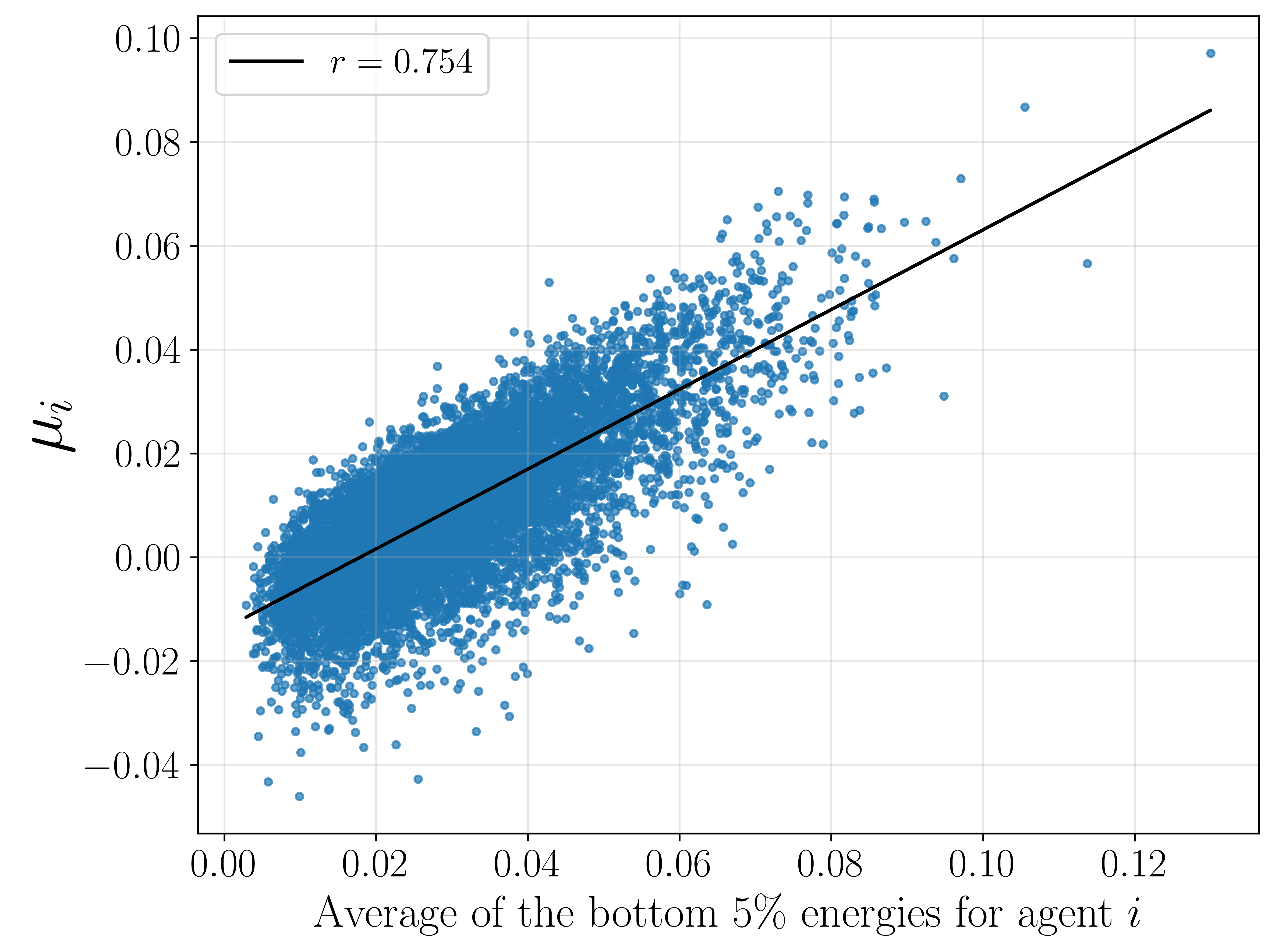}
    \label{fig:mu_a}
\end{subfigure}

\begin{subfigure}{\textwidth}
    \caption{$k=2$}  
    \centering
    \includegraphics[scale=0.5]{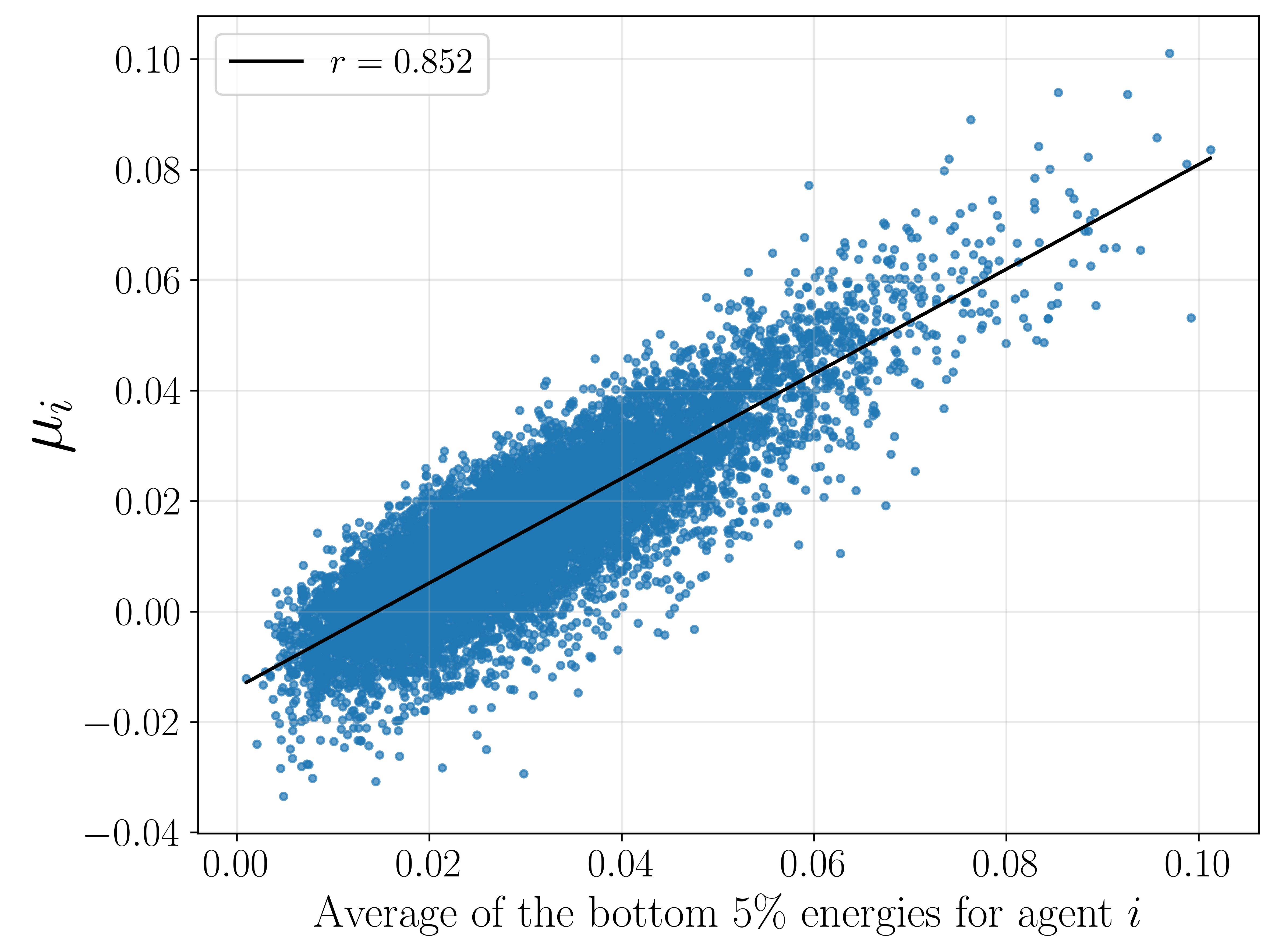}
    \label{fig:mu_b}
\end{subfigure}

\caption{Chemical potential of agent $i$ versus the average energy computed over the bottom $5\%$ of energies for agent $i$, i.e., $\langle E_{ij} \rangle_{\text{bottom } 5\%\, j}$. The points are taken from 5000 realizations with $N = 100$ and $\beta = 600$, with $k = 1$ in panel (a) and $k = 2$ in panel (b). In general, the higher the chemical potential of an agent, the less inclined the agent is to form connections in the grand-canonical framework. In the legends, $r$ is the Pearson correlation coefficient.}
\label{fig:mu}
\end{figure}

\begin{figure}[p]
\centering

\begin{subfigure}{\textwidth}
    \caption{$k=1$}  
    \centering
    \includegraphics[scale=1]{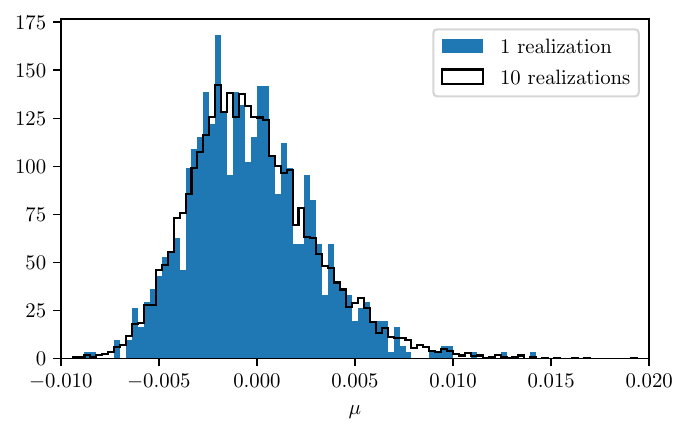}
    \label{fig:mu_a2}
\end{subfigure}

\begin{subfigure}{\textwidth}
    \caption{$k=2$}  
    \centering
    \includegraphics[scale=1]{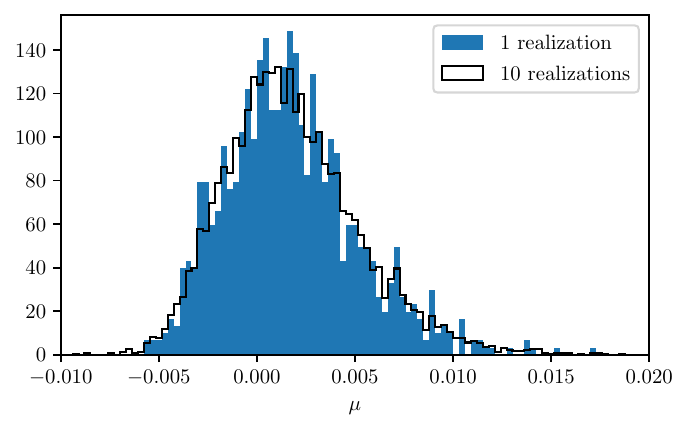}
    \label{fig:mu_b2}
\end{subfigure}

\caption{Distribution of the variables $\{\mu_i\}_{i=1,..,N}$ for $k=1$ (top panel) and $k=2$ (bottom panel) with $N=100$ and $\beta=600$. The blue bars show a single realization while the black lines correspond to an average over 10 realizations. One realization of the system can describe the probability distribution, suggesting a self averaging behavior.}
\label{fig:mus}
\end{figure}

In contrast to saddle-point-based derivations, where the variables \(\mu_i\) appear as auxiliary variables enforcing hard constraints, in our framework the \(\mu_i\) naturally represent chemical potentials, acting as local order parameters associated with each agent and can be interpreted as their propensity to form connections. This is supported by Figure \ref{fig:mu}, showing that the higher the chemical potential associated with a node $i$ at a given temperature, the larger its lowest interaction energies $E_{ij}$. This indicates that such a node
is less prone to form connections or accept compromises, providing a direct interpretation of the variable $\mu_i$. Furthermore, as illustrated by their distribution in Fig.~\ref{fig:mus}, the chemical potentials exhibit a self-averaging property: a single realization of the disorder (i.e., of the energy matrix) appears sufficient to describe the behavior of the entire system. This observation hints at a potential connection between our chemical potentials and the replica-symmetry ansatz introduced in \cite{mezard1985replicas} for solving random optimization problems. While this connection is speculative at this stage, it suggests an intriguing direction for future investigation.

Finally, our approach yields an analytically tractable partition function in Eq.~\eqref{partition2}, from which physically meaningful quantities can be directly derived. For instance, it is immediate from Eq.~\eqref{condition} that the summands in Eq.~\eqref{chemical_potentials} define a finite-temperature agent network, and that the probability of a link between any two agents is given by Eq.~\eqref{probV}, where \(V_{ij}\) explicitly appears in the constraint equations.

\section{Conclusions}
In this paper, we reformulated random optimization problems as an arrangement model and developed a polynomial-time procedure to solve the matching problem, the minimum-weight
k-factor problem, and, in the thermodynamic limit, the Traveling Salesman Problem. We also derived an analytical expression for the average ground-state energy.
Our method is based on a grand-canonical approach, in which each node in the network is assigned a chemical potential. This relaxes the strict requirement that every agent must occupy exactly one seat in the arrangement configuration, yielding an analytically tractable partition function. In this way, our framework avoids the need for more elaborate methods such as the replica or cavity approaches commonly used in statistical physics and combinatorial optimization. It would however be interesting to explore in future work how these chemical potentials relate to the replica-symmetry ansatz in~\cite{mezard1986replica}. 
We have described how the key ingredient enabling this relaxation from fixed constraints to average constraints is the double-stochasticity property of the matrix $V_{ij}$ for any temperature. This guarantees that the solution converges to the exact optimum of both the matching problem and the minimum-weight k-factor problem. For the TSP, the presence of the additional constraint of forming a single Hamiltonian cycle breaks double stochasticity; nevertheless, it is known that in thermodynamic limit only local properties matter to compute the energy of the TSP solution, and this global constraint becomes negligible. As a consequence, the TSP becomes equivalent to the minimum-weight 2-factor problem at the level of ground-state energy. In principle, our framework has a broader range of applicability than the three problems considered in this paper. Indeed, we have shown how different constraint structures map the Hamiltonian in Eq.~\eqref{H} onto that of a wider class of constrained optimization problems. However, in the models considered in this work the constraint topology is relatively simple, i.e., the adjacency matrix $\mathbf{A}$ is regular. The question remains as to whether, for more complex topologies, properties analogous to double stochasticity exist that ensure convergence to the true optimum, provided the constraints can be written in a tractable form. Alternatively, as in the TSP case, it may be possible to show that such complex constraints become irrelevant in the thermodynamic limit. Thus a potential future research direction is to investigate the applicability and limitations of the grand-canonical framework in these broader settings. To illustrate potential issues, Appendix D discusses the challenges that arise when applying our method to a highly frustrated optimization problem, the Sherrington–Kirkpatrick model~\cite{panchenko2013sherrington}.

Overall, the simplicity of our method, and the way it complements existing approaches, offers new insights into the connection between statistical physics and combinatorial optimization. 

\clearpage
\appendix
\setcounter{figure}{0}
\renewcommand{\thefigure}{B.\arabic{figure}}

\section{Numerical implementation}
The central result of our approach is the nonlinear system of \(N\) equations described in Eq.~\ref{chemical_potentials}. For a given inverse temperature \(\beta\) and instance of the energy matrix \(\epsilon\), we solve this system using the Newton–Krylov iterative method, which, in contrast to standard Newton-like methods for solving nonlinear equations, is Jacobian-free~\cite{martinez2000practical}. That is, it does not require the explicit computation or inversion of the Jacobian matrix at each iteration, which can be computationally costly. The method achieves quadratic convergence when the initialization of the solution vector (the initial values of \(\mu_i\)) is sufficiently close to the true solution.

To compute the minimum energy defined in Eq.~\ref{eq:min_energy}, it is necessary to consider large values of \(\beta\), which make the system of equations highly nonlinear and can negatively affect the convergence of the iterative procedure. We thus adopt an annealing strategy, as already proposed in~\cite{kosowsky1994invisible} for solving a similar system of equations. In this approach, the temperature is gradually decreased over time steps, and the solution vector \(\mu_i(t)\) found at step \(t\) for inverse temperature \(\beta(t)\) is used as the initialization for solving the system at step \(t+1\). This ensures that the initial guess at each step remains close to the corresponding solution, thereby improving the convergence of the Newton–Krylov method.

At each time step, the energy is computed as:
\begin{equation}\label{umin}
U_{\beta(t)} = \frac{1}{2} \sum_{ij} \frac{E_{ij}}{1 + e^{\frac{\beta}{2}\left(E_{ij} - \mu_i(t) - \mu_j(t)\right)}}.
\end{equation}
The annealing procedure is terminated when the energy difference between two successive steps satisfies \(|U_{\beta(t+1)} - U_{\beta(t)}| < \varsigma\), where \(\varsigma\) is typically set to \(10^{-5}\). The resulting value \(U_{\min}\) is taken as the minimum energy of our arrangement model for the specific instance of the energy matrix \(\epsilon\). Although we do not provide a theoretical bound (as exists for the Hungarian algorithm \cite{fredman1987fibonacci}), the computational complexity of this process is empirically observed to be of order \(\mathcal{O}(N^2)\). This is mainly due to the cost of solving the system of equations, as both the number of Newton–Krylov iterations and the number of annealing steps needed to reach \(U_{\min}\) are typically much smaller than \(N\).

To obtain the the configurational average of the minimum energy, the entire annealing procedure is repeated across multiple realizations of the energy matrix \(\epsilon\). The entire process is schematically described below.

\begin{algorithm}
\caption{Annealing Algorithm for Computing \(\langle U_{\min} \rangle\)}
\begin{algorithmic}
\Require Number of agents \(N\), number of realizations \(R\), convergence threshold \(\varsigma\)
\Ensure Estimated configurational average energy \(\langle U_{\min} \rangle\)

\State \textbf{Initialize:} \(\texttt{U\_collect} \gets 0\)
\For{\(r = 1\) to \(R\)}
    \State Generate energy matrix \(\epsilon^{(r)}\)
    \State Initialize \(\mu_i(0) = 0\), \(\beta(0)\), \(t \gets 0\)
    \While{not converged}
        \State Solve Eq.~\ref{chemical_potentials} via Newton–Krylov with \(\mu(t)\) and \(\beta(t)\), get \(\mu(t+1)\)
        \State Compute \(U_{\beta(t)}\) using Eq.~\ref{umin}
        \If{\(\left|U_{\beta(t+1)} - U_{\beta(t)}\right| < \varsigma\)}
            \State \(\texttt{U\_collect} \gets \texttt{U\_collect} + U_{\beta(t+1)}\)
            \State \textbf{break}
        \EndIf
        \State Increase \(\beta(t+1)\), set \(t \gets t+1\)
    \EndWhile
\EndFor
\State \Return \(\langle U_{\min} \rangle = \texttt{U\_collect} / R\)
\end{algorithmic}
\end{algorithm}

\section{Approaching Matching and TSP solutions with $\beta$}
In this appendix, we present simple examples of cost-matrix instances for both the
matching problem ($k = 1$) and the TSP ($k = 2$). In the first case, we show how the grand-canonical solution converges to the exact optimal configuration when lowering the temperature. In the second case,
we provide a heuristic greedy algorithm that uses the link probabilities of
Eq.~\eqref{probV} to approximate the solution of the TSP.

\subsection*{Exact convergence to the Matching problem solution}

As we mentioned in the main text, it is possible to show that in the case of the matching problem our procedure guarantees that the expected value of $V_{ij}$, or equivalently, the probability that a link exists between agent $i$ and agent $j$ obtained in Eq.~\eqref{probV}, converges to the configuration of the true ground state, i.e.\ the configuration that minimizes the matching energy. This follows from the properties of $V_{ij}$ arising from its double-stochastic structure. 

\begin{figure}[h!]
    \centering
    \includegraphics[width=1\linewidth]{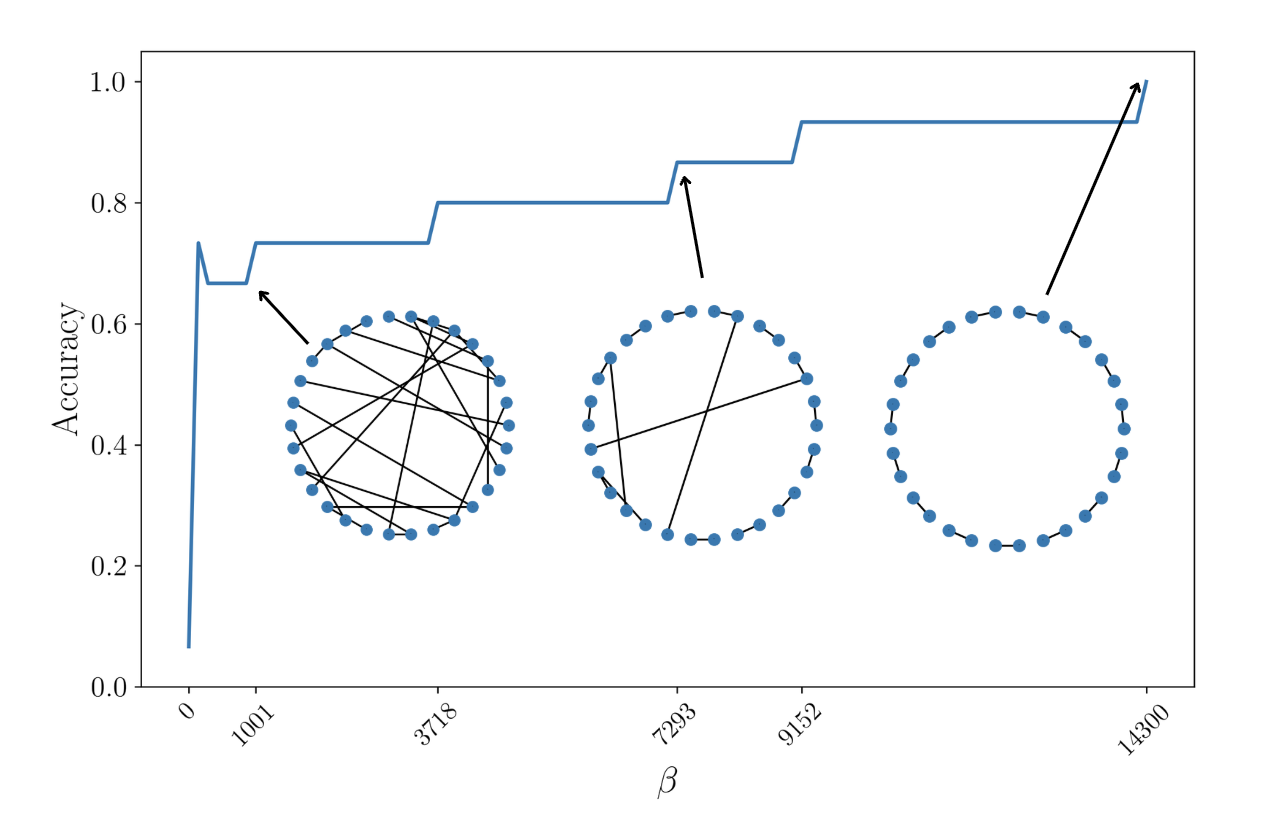}
    \caption{Number of correctly identified links (accuracy) as a function of the inverse temperature $\beta$. We consider a system with $N = 30$ and a randomly generated symmetric cost matrix. For each value of $\beta$, the system in Eq.~\eqref{chemical_potentials} is solved and the resulting $\mu_i$ are inserted into Eq.~\eqref{probV} to obtain the link–probability matrix $V_{ij}$. The network is then constructed by associating each node $i$ with the node $j$ that maximizes the corresponding row $V_{ij}$. The three networks shown illustrate the extracted configurations at three different temperatures. The last graph on the right is the optimal matching.}
    \label{exmatching}
\end{figure}

Here we illustrate this with a concrete example. In particular, in Fig.~\ref{exmatching} we generated a random symmetric cost matrix with $N=30$ agents and observed how the solution changes with the temperature. For a given $\beta$, we solve the system in Eq.~\eqref{chemical_potentials}, determine the corresponding $\mu_i$, and plug them into Eq.~\eqref{probV} to obtain the link probabilities $V_{ij}$. Given the obtained probability matrix $V_{ij}$, among the various possible ways to sample a configuration we chose, for the purpose of the figure, to assign to each node $i$ the node $j$ for which $V_{ij}$ is maximal.  
The figure shows how the accuracy, defined as the number of correctly identified links with respect to the true optimal matching configuration, improves as the temperature decreases. For small $\beta$, the matrix $V_{ij}$ is essentially uniform.  
With the drawing rule adopted here, each node is guaranteed to have at least one link.  
Thus, for low $\beta$, it is very likely that a node will have more than one link, since several nodes may have the same $j$ as their maximal $V_{ij}$. As $\beta$ increases, we know that $V_{ij}$ tends to a $\theta$-function, taking values only in $\{0,1\}$, and as shown in the figure, our procedure eventually yields a perfect matching corresponding to the optimal solution.

As a side note, another natural way to draw the network would have been to sample links directly from the probabilities $V_{ij}$.  
However, in that case one would typically need to reach larger values of $\beta$ before observing almost with certainty the correct optimal configuration.

\subsection*{A "grand-canonical heuristic" for the Traveling Salesman problem}

The case of the TSP ($k = 2$) has an additional global constraint: the final configuration must be a single Hamiltonian cycle. This implies that our procedure is exact only for the value of the ground–state energy in the limit $N \to \infty$. In practice, we neglect this global constraint, and when we take $\beta \to \infty$ the configuration obtained from the values of $V_{ij}$ is not a single cycle but a set of smaller disjoint cycles. Such a configuration satisfies the local constraint $k = 2$ but not the global requirement of having a unique cycle, and therefore it corresponds to the true solution of the minimum–weight 2–factor problem. Here we propose a greedy algorithm which, starting from this $k=2$ solution composed of many smaller cycles and exploiting the nonzero values of $V_{ij}$ for finite $\beta$, efficiently tends toward the actual TSP solution.

The first observation is that, for any energy matrix, the solution of the minimum–weight 2–factor problem has energy less than or equal to the energy of the TSP solution. This is because the 2–factor problem has fewer constraints. As already mentioned, the solution of the minimum weight 2–factor problem is a set of disjoint cycles. The main idea is therefore to connect these small cycles into a single cycle in an optimal way, so as to approach as closely as possible the TSP solution. To decide which link to remove from the small closed cycles and in what order to connect these cycles, we use the values of $V_{ij}$ at finite $\beta$ in a greedy manner.

The procedure is as follows. First, we compute the optimal 2–factor solution in the limit $\beta \to \infty$. We then select at random one of the cycles in this solution and cut a randomly chosen link. After that, we increase the temperature to obtain finite values of $V_{ij}$. One extremity of the cut cycle is then connected to the node in another cycle for which $V_{ij}$ is maximal. At this point we remove the redundant link in the second cycle, thereby forming an arc larger than the previous one. We repeat this process until no small cycles remain, at which point we close the two extremities of the final arc to obtain a single cycle. This merging procedure requires $O(N^2)$ steps. If one wants to further lower the energy of the resulting solution, the entire procedure can be repeated by choosing different initial cycles and initial links, and selecting the configuration with the lowest energy. In the worst case this requires at most $2^{N/3}$ additional attempts, since $N/3$ is the maximum possible number of cycles in a 2–factor. If the average number of cycles in random instances of the optimal 2–factor scales as $\log N$ (as in the case of a uniform random 2–regular graph \cite{bollobas2011random}), then the number of steps scales as $2^{\log N}$. The algorithm is detailed below.

\begin{algorithm}\label{tsp_algorithm}
\caption{Gran-Canonical Greedy Procedure the TSP Solution}
\begin{algorithmic}
\Require Energy matrix $\epsilon$, inverse temperature $\beta$
\Ensure Approximate Hamiltonian cycle

\State Compute the minimum weight 2--factor solution in the limit $\beta \to \infty$, obtaining a set of $M$ cycles $\mathcal{C}_0$
\State $\texttt{BestCycle} \gets$ None
\State $\texttt{BestEnergy} \gets +\infty$

\For{$m = 1$ to $M$} 
    \State $\mathcal{C} \gets$ copy of $\mathcal{C}_0$
    \State Choose the $m$-th cycle $C$ in $\mathcal{C}$ and cut a random link, obtaining an open arc $A$
    \State Increase temperature to obtain finite $V_{ij}$ values

    \While{$|\mathcal{C}| > 1$}
        \State Let $i$ be an extremity of the current open arc $A$
        \State Find node $j$ in a different cycle such that $V_{ij}$ is maximal
        \State Connect $i$ to $j$
        \State Remove the redundant link in the cycle containing $j$
        \State Remove the merged cycle from $\mathcal{C}$
    \EndWhile
    
    \State Close the two extremities of $A$ to form a single Hamiltonian cycle $H$
    \State Compute the total energy $U(H)$

    \If{$U(H) < \texttt{BestEnergy}$}
        \State $\texttt{BestEnergy} \gets U(H)$
        \State $\texttt{BestCycle} \gets H$
    \EndIf
\EndFor

\State \Return $\texttt{BestCycle}$

\end{algorithmic}
\end{algorithm}

We tested this algorithm on small, standard benchmark cost matrices. In particular, we considered the instances \texttt{burma14}, \texttt{ulysses16}, and \texttt{ulysses22}. These benchmarks are part of the TSPLIB collection \cite{reinelt1995tsplib95}, a publicly available library of TSP instances used for validating new algorithms. The instance \texttt{burma14} is a 14–node TSP defined over geographical coordinates from Burma, while \texttt{ulysses16} and \texttt{ulysses22} contain respectively 16 and 22 locations extracted from the trajectory of the mythical voyage of Ulysses. Their optimal TSP tour lengths are known to be $3323$ for \texttt{burma14}, $6859$ for \texttt{ulysses16}, and $7013$ for \texttt{ulysses22}, as documented in the TSPLIB database. Remarkably, in all these cases our algorithm converged to the exact TSP optimal solution. Note that a naive greedy procedure, where in the merging phase one replaces the $V_{ij}$ values with the bare cost matrix entries $E_{ij}$, does not recover the optimal solution. 

Nevertheless, several open questions remain. First, it is not yet clear which value of $\beta$ is sufficient to obtain $V_{ij}$ values that best guide the merging of the small cycles toward the true TSP tour (in the three examples above, the higher the number of nodes the higher the needed temperature). From a theoretical point of view, we still do not know to what extent the finite–temperature $V_{ij}$, which are designed to solve the minimum weight 2--factor problem, can also assist in identifying the minimum–weight Hamiltonian cycle. Intuitively, lowering the temperature drives $V_{ij}$ toward one of the vertices of the convex polytope of doubly stochastic matrices, whose vertices correspond to permutation matrices. When $\beta = 0$, the matrix $V_{ij}$ is far from any vertex, whereas increasing $\beta$ moves $V_{ij}$ closer to the vertex associated with the minimum weight 2--factor solution. However, if $\beta$ is too large, $V_{ij}$ becomes too close to this 2--factor vertex and correspondingly too far from the vertex associated with the TSP solution. Hence, a clear characterization of the optimal range of $\beta$ is an open problem.

\section{Saddle-point derivation of Eq.~\eqref{chemical_potentials}}

In this appendix we present an alternative derivation of the system of
equations~\eqref{chemical_potentials}, in which the hard constraints are relaxed
through a saddle-point approximation.

Given the energy $U(V)$ associated with a configuration $V$ of the agent network,
the probability of observing $V$ at inverse temperature $\beta$ in statistical
physics is given by the Gibbs distribution,
\begin{equation}
    P(V) = \frac{e^{-\beta U(V)}}{Z},
\end{equation}
where $Z$ is the partition function,
\begin{equation}
    Z = \int_{V} e^{-\beta U(V)}\, dV.
\end{equation}

In our discrete, monopartite setting, the variables $V_{ij}$ take binary values
$0$ or $1$. Incorporating the degree constraint in
Eq.~\eqref{condition}, the partition function becomes
\begin{equation}
    Z = \sum_{V_{ij}\in\{0,1\}}
        e^{-\frac{\beta}{2}\sum_{ij} E_{ij}V_{ij}}
        \prod_{i} \delta\!\left(\sum_{j} V_{ij} - k \right).
\end{equation}

Using the Fourier representation of the $\delta$-functions, we introduce the variables $\mu_j$, which will play the same role as the chemical potentials appearing in the grand-canonical derivation:
\begin{equation}
    Z = \int_{-\infty}^{+\infty} \prod_{j} d\mu_j\, 
        e^{\beta \sum_{j} i\mu_j}
        \sum_{V_{ij}\in\{0,1\}}
        e^{-\beta \sum_{ij} V_{ij}\,[E_{ij} + i\mu_i + i\mu_j]}.
\end{equation}

The sum over the binary variables factorizes, yielding
\begin{equation}
    Z = \int_{-\infty}^{+\infty} \prod_{j} d\mu_j\,
        \exp\!\left(
            -\beta \sum_{j} i\mu_j
            - \frac{1}{\beta} \sum_{ij}
            \ln\!\big[1 + e^{-\beta(E_{ij} + i\mu_i + i\mu_j)}\big]
        \right).
\end{equation}

Since the combinations $i\mu_j$ must ultimately be real, we perform the change
of variables $\mu_j \rightarrow i\mu_j$. The resulting expression for $Z$ is not
analytically tractable, and therefore one has to apply the saddle-point
approximation, where the assumption is that, for large $N$, the dominant contribution to the
integral arises from the stationary point of the exponent. Taking the derivative of the exponent with respect to each $\mu_j$ and setting it to zero yields the
saddle-point conditions. One can verify that these conditions reproduce exactly the system of equations in~\eqref{chemical_potentials}.

\section{Generalization}

In this paper we introduced the grand-canonical ensemble and directly wrote the partition function, obtaining iterative equations for a set of variables $\{\mu_i\}_{i=1,\dots,N}$ that define the statistics of the problem.  
With general constraints, such a partition function can be derived as follows. Consider the minimization problem under constraints:
\begin{align}
    &\min_{x \in \Omega_N} H(x), \\
    &C_i(x) = 0, \qquad i = 1,\dots,M,
\end{align}
where $\Omega_N$ denotes the set of all the possible configurations $x$ of the system, $H(x)$ is the Hamiltonian to be minimized, and $C_i(x)$ are the constraints on $x$.  
Typically, $x$ is an $N$-dimensional vector or, in our case, a matrix $A_{ij}$.

We now introduce the following Lagrangian relaxation, corresponding to the minimization of a Lagrangian with the addition of an entropic term:
\begin{equation}
\begin{aligned}
L\big(p(x),\beta,\{\mu_i\}\big)
    &= \sum_{x \in \Omega_N} p(x)\, H(x)
    + \beta \sum_{x \in \Omega_N} p(x)\, \log p(x)+ \\
    &\quad + \sum_{i=1}^M \mu_i 
        \left( \sum_{x \in \Omega_N} p(x)\, C_i(x) \right).
\end{aligned}
\end{equation}
over the variables \(\beta\), \(\mu_i\), and the probability distribution of the configurations \(p(x)\). The variational problem yields the following equations:
\begin{equation}
    \frac{\delta L}{\delta p(x)} = 0
    \quad \Rightarrow \quad
    p(x) = \frac{1}{Z} \exp\!\left[-\beta H(x) + \sum_{i=1}^M \mu_i C_i(x)\right],
\end{equation}
with the partition function
\begin{equation}
    Z(\beta,\mu_i) = \sum_{x \in \Omega_N} 
    \exp\!\left[-\beta H(x) + \sum_{i=1}^M \mu_i C_i(x)\right].
\end{equation}

The stationarity conditions with respect to $\beta$ and $\mu_i$ give:
\begin{align}
    \frac{\partial L}{\partial \beta} &= \sum_{x \in \Omega_N} p(x) \log p(x) = 0
    \quad \Rightarrow \quad
    \sum_{x \in \Omega_N} H(x)\, p(x)
    = -\frac{1}{\beta} \log Z, \\
    \frac{\partial L}{\partial \mu_i} &= \sum_{x \in \Omega_N} p(x) C_i(x) = 0
    \quad \Rightarrow \quad
    \frac{\partial}{\partial \mu_i} \log Z = 0.
\end{align}

By introducing the above Lagrangian, what we have essentially done is to relax the constraints from exact conditions to average ones: instead of requiring $C_i(x) = 0$ exactly, we now impose $\langle C_i(x) \rangle = 0$. In this way, for finite temperature we obtain a lower bound on the true minimum energy of the problem:
\begin{equation}
   \sum_{x \in \Omega_N} p(x)\, H(x) =F(\beta, \mu_i) \leq \min_{x \in \Omega_N} H(x).
\end{equation}
And for $\beta \to \infty$ they coincide.

Hence, the partition function $Z(\beta,\mu_i)$ solves the minimization process. If $Z(\beta,\mu_i)$ is analytically treatable, as in the optimization problems we have studied in this paper, then all the other conditions on $\mu_i$ and $\beta$ can be solved through a simple derivative. Indeed, in our case, the first equation corresponds to the ensemble average energy in \eqref{eq:min_energy} that, in turn, is the free energy of the system $F(\beta, \mu_i)$, and the second one to the constraint equation in \eqref{chemical_potentials}.

In particular, for the matching problem, where linear constraints can be written as in D.2, this approach leads to the exact solution in the zero-temperature limit. For the TSP, we make the additional approximation of keeping only linear terms in the energy and in the constraints, implicitly assuming that only local properties are important.  
This is justified only in the large-$N$ limit and only when one is interested in the average ground-state energy, due to the result discussed in \cite{frieze2004random}. Regarding the chemical potentials, this linearity assumption makes them independent, thereby avoiding the issues that arise when the chemical potentials are correlated (see, for example, \cite{baybusinov2024nonrandom}).


Note that in our case the properties of $\mathbf{A}$ were regular, meaning that each $k_i$ was equal to a constant $k$, but in many applications they can be much more complex.  
As an example, we now show the equation obtained by applying the same grand-canonical framework to a classical model in the physics of disordered systems, namely the Sherrington--Kirkpatrick (SK) model introduced to study spin glasses \cite{panchenko2013sherrington}.  
For this model, for which we refer the reader to standard textbooks on disordered systems and spin glasses, such as \cite{nishimori2001statistical}, the network matrix is given by $A_{ij} = \sigma_i \sigma_j$, where $\sigma_i = \pm 1$ denotes the spin at site $i$, and the constraint reads
\[
k_i = \sigma_i \sum_{k \neq i} \sigma_k.
\]
For a given realization of the spins, following the same procedure as above, one can show that at a given temperature the analogue of Eq.~\eqref{chemical_potentials} becomes
\begin{equation}
    \sigma_i \sum_{k \neq i} \sigma_k
    = \sum_{j \neq i} \tanh\!\bigl( \beta(\mu_i + \mu_j - \epsilon_{ij}) \bigr).
\end{equation}

From this point, one must determine the $\mu_i$ in order to compute the probability of each configuration and then average the energy over all possible degrees, in analogy with Eqs.~\eqref{eq:min_energy} and \eqref{eq_media}:
\begin{align}
    P(A_{ij}) &= 
    \frac{
        e^{-\beta (\mu_i + \mu_j - \epsilon_{ij}) A_{ij}}
    }{
        2 \cosh\!\bigl( \beta(\mu_i + \mu_j - \epsilon_{ij}) \bigr)
    },\\[4pt]
    \bigl\langle H_{\min} \bigr\rangle_{\epsilon}
        &= \binom{N}{2} \,
           \lim_{\beta \to \infty}
           \bigl\langle \epsilon_{ij} A_{ij} \bigr\rangle_{\epsilon}.
\end{align}

The difficulty is that Eq.~D.9 is not in closed form, since the variables $\mu_i$ depend on the spins $\sigma_i$ themselves.  
Thus one must either rely on suitable approximations or implement convergent numerical algorithms.  
This discussion is not meant to provide a full solution, but rather to illustrate the potential issues that arise in highly frustrated systems and to highlight possible directions for future research.

\section*{acknowledgments}
This work was partially supported by the Swiss National Science Foundation (SNF). In particular, E.M.F. acknowledges the SNF Postdoc.Mobility grant No. $P500PT\_225431$.

\bibliographystyle{elsarticle-num-names} 
\bibliography{cas-refs.bib}



\end{document}